\documentclass{elsart3p}
\usepackage{amsmath}
\usepackage{graphicx}
\usepackage{epsf}
\usepackage{graphics}
\usepackage[latin1]{inputenc}
\newcommand{\be}{\begin{equation}}
\newcommand{\ee}{\end{equation}}
\newcommand{\bea}{\begin{eqnarray}}
\newcommand{\eea}{\end{eqnarray}}

\begin{document}

\begin{frontmatter}

\title{Fermionic Contributions to the Free Energy of Noncommutative
Quantum Electrodynamics at High Temperature}

\author{F. T. Brandt, J. Frenkel and  C. Muramoto}
\address{Instituto de Física, Universidade de São Paulo, 05508-090, São Paulo, SP, BRAZIL}

\begin{abstract}
We consider the fermionic contributions to the free energy of
noncommutative QED at finite temperature $T$.
This analysis extends the main results of our previous
investigation where we have considered the pure bosonic sector of the theory.
For large values of $\theta T^2$ 
($\theta$ is the magnitude of the noncommutative parameters) the fermionic contributions decrease
the value of the critical temperature, above which there occurs a thermodynamic instability.
\end{abstract}


\end{frontmatter}

\section{Introduction}
The formulation of Quantum Electrodynamics in noncommutative space
(NCQED) \cite{Douglas:2001ba,Szabo:2001kg,Hayakawa:1999yt} opens the interesting possibility of new
dynamics for the gauge fields, including self-interactions, without the introduction of
additional internal degrees of freedom. The gauge field self-interactions of this theory contain
trigonometric factors like $\sin(p_\mu\,\theta^{\mu\nu} q_\nu)$ where $p_\mu$ and
$q_\nu$ are the momenta and $\theta^{\mu\nu}$
are the noncommutative parameters introduced via the commutation relation
of the coordinates
\be
[x^\mu,x^\nu]_\star = x^\mu\star x^\nu - x^\nu\star x^\mu = i\theta^{\mu\nu},
\ee
where, in general, the Grönewold-Moyal $\star$-product \cite{Douglas:2001ba,Szabo:2001kg}
between two functions $f(x)$ and $g(x)$ is given by
\begin{eqnarray}\label{moyal1}
f(x)\star g(x) = f(x)~\exp\left(\frac{i}{2}\,\overleftarrow{\partial_\mu}\,
\theta^{\mu\nu}\,\overrightarrow{\partial_\nu}\right)~g(x) ,
\end{eqnarray}
In the limit when the coordinates can be considered commutative,
$\theta^{\mu\nu}\rightarrow 0$, all the self-interactions vanish and
the usual theory of non-interacting photons is recovered.

When we consider a system at finite temperature $T$\cite{kapusta:book89,lebellac:book96,das:book97},
the thermal Green functions may became dependent on the quantity
$\tau \equiv \theta\,T^2$, were
$\theta$ is the typical magnitude of the components $\theta^{\mu\nu}$.
In general the dependence on $\tau$ may be very involved. However, 
for asymptotic values of $\tau$ one can have
a simpler understanding of the behavior of thermal Green functions.
It is also possible that other mass scales may be present, so that one could
have quantities like $\theta p T$, where $p$ represents the magnitude
of some external momenta in a given Green function. 
There is however an important quantity, namely the {\it free-energy}, 
which can only depend on $\tau$. This happens because
the diagrams which contribute to the free-energy do not have external legs.

Previous investigations on this subject \cite{Arcioni:1999hw}
have revealed interesting
properties already at the lowest non-trivial order (two-loop order). 
Corrections to the free energy at higher orders than two-loops, in thermal field
theories, require for a self-consistent treatment to take into
consideration an infinite series of
diagrams. This happens because the fields acquire, through their interactions,
an effective thermal mass. This so-called {\it plasmon effect} is
known to occur for instance in non-abelian gauge theories like the
Yang-Mills theory at high temperature \cite{kapusta:book89,lebellac:book96,das:book97}.
Inasmuch NCQED is a theory of self-interacting gauge fields, it
is natural to consider the possibility that similar non-perturbative effects
may arise when we take into account all the higher order corrections
to the free energy. In this case, only the leading hard thermal contributions are
relevant. Consequently, other interesting phenomena such as the 
UV/IR mixing \cite{Minwalla:1999px}, which are important at $T=0$, do not arise in the present context
since the temperature does provide an UV regularization of the thermal graphs.

In a recent paper the free-energy of thermal NCQED (TNCQED)
has been analyzed by taking into account all the contributions
which arise from the pure bosonic sector of TNCQED \cite{Brandt:2006bf}.
In the regime when $\tau \gg 1$, we have summed the {\it ring-diagrams}, 
which consist of an arbitrary number of
self-energy insertions and yield the leading contributions.
A remarkable property revealed by this analysis
was the emergence of a thermodynamic instability above the critical
temperature
\be\label{tc1}
T_c = \sqrt{\frac{1}{e\theta }\frac{3}{2\pi}\sqrt{10}} \approx 1.229 \, \sqrt{\frac{1}{e\theta }}
\ee
where $e$ is the coupling constant.

In the present paper we consider the possibility that the critical temperature
$T_c$ may be modified by the inclusion of fermions. If we consider only
Dirac fermions in the fundamental representation,
there would be no modification of $T_c$, because, as we have argued in \cite{Brandt:2006bf},
the contribution from fermion loops are the same as in the commutative
theory and the existence of $T_c$ is a consequence of the
noncommutative magnetic mode, which is absent in the commutative theory.
However, we may additionally include fermions in the ``adjoint'' representation
so that the fermionic action, for massless fermions, is given by
\bea\label{lagraferm}
S_{\rm ferm} &=& \int {\rm d}^4 x \left[
 \bar\psi\star i \left(\slash\!\!\! \partial \psi -i e\,\gamma_\mu A^\mu\star\psi\right)
\right. \\ \nonumber
\qquad & + & \left.
 \bar\chi\star i \left(\slash\!\!\! \partial \chi -ie\,\gamma_\mu[A^\mu,\chi]_\star\right) 
\right],
\eea
where $A^\mu$ is the gauge field. The fields $\psi$ and $\chi$ are the
fundamental and ``adjoint'' fermions, which transform
respectively as \cite{Hayakawa:1999yt}
\be
\psi \rightarrow {\rm e}^{i\alpha(x)}\star \psi
\;\;\mbox{and}\;\;
\chi \rightarrow {\rm e}^{i\alpha(x)}\star \chi \star {\rm e}^{-i\alpha(x)},
\ee
where $\alpha(x)$ is the parameter of the U(1) group.
In the limit when $\theta^{\mu\nu}\rightarrow 0$ the adjoint fermions
decouple and we recover the usual electron interaction of commutative QED.

In section {\bf 2} we describe the method employed to
identify the leading contributions to the free-energy in terms of ring
diagrams. Then, in section {\bf 3}, we present the result for the
free-energy, in a form which allows us to derive the critical
temperature, which turns out to be smaller than the one in the pure
bosonic sector. The instability which occurs above the critical
temperature may be understood, as discussed in section {\bf 4}, in
terms of the behavior of the noncommutative thermal magnetic mass.
Many technical details, which are relevant in the context of these
calculations, can be found in our previous paper \cite{Brandt:2006bf}.

\section{Basic Approach}

Let us review the main steps which lead us to find the critical
temperature $T_c$ in \cite{Brandt:2006bf}. First we have to consider a
non-perturbative 
(a very comprehensive analysis of the nonpertubative regime has been
carried out in  \cite{Bietenholz:2006cz} using the lattice formulation)
contribution to the free-energy which is generated by the {\it ring diagrams},
as follows
\be\label{eq1}
\tilde\Omega^{\rm r}(T,\theta) =
-\frac{1}{2}\left[
\frac{1}{2}\begin{array}{c}\includegraphics[scale=0.25]{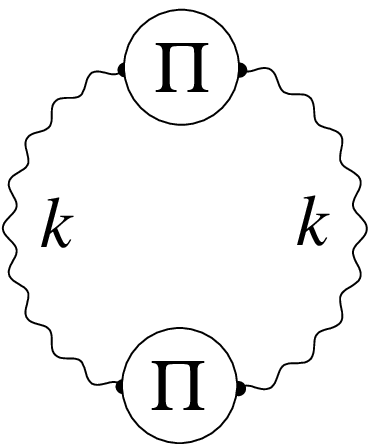}\end{array}
-\frac{1}{3}\begin{array}{c}\includegraphics[scale=0.25]{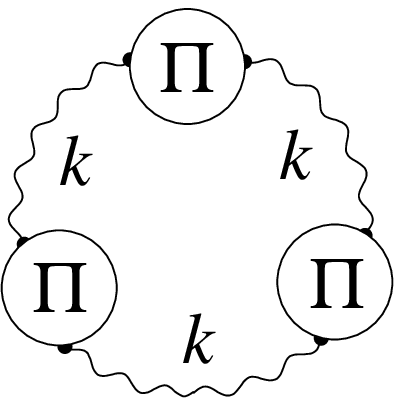}\end{array}
+\cdots 
\right].
\ee
The quantity $\Pi$ stands for all the contributions to the one-loop photon self-energy.
In \cite{Brandt:2006bf} we have shown that at any given order in the
loop expansion, Eq. \eqref{eq1} gives the leading contribution in the limit $\tau\gg 1$. 
Therefore, it is implicit in \eqref{eq1} that we are considering 
only the {\it zero mode} contribution of the photon propagator (denoted by wavy lines), 
so that $k^2 = -|\vec k|^2$. The reason for this is because each
photon propagator in \eqref{eq1} introduces a factor
\be
\frac{1}{k^2} = -\frac{1}{(2\pi n)^2 + |\vec k|^2/\tau^2},
\ee
so that the zero mode, $n=0$, yields the following leading contribution for large
values of $\tau$
\be
\frac{1}{k^2} \approx -\frac{\tau^2}{|\vec k|^2}
\ee
(here and in all the expressions which follows 
we have performed the rescaling $(k_4,\vec k)\rightarrow T(k_4,\vec k/\tau)$
as well as $(p_4,\vec p)\rightarrow T(p_4,\vec p)$ where $p^\mu$ stands
for the internal momentum of $\Pi_{\mu\nu}$).
On the other hand, possible contributions to \eqref{eq1} with
fermionic rings (photon lines replaced by fermion lines) would be
sub-leading, because the inverse of the fermionic propagator is nonzero for $n=0$ and $|\vec k|=0$
(the Matsubara frequencies are all multiples of odd integers).

\begin{center}
\begin{figure} 
\[
\begin{array}{cc}
\begin{array}{c}\includegraphics[scale=0.45]{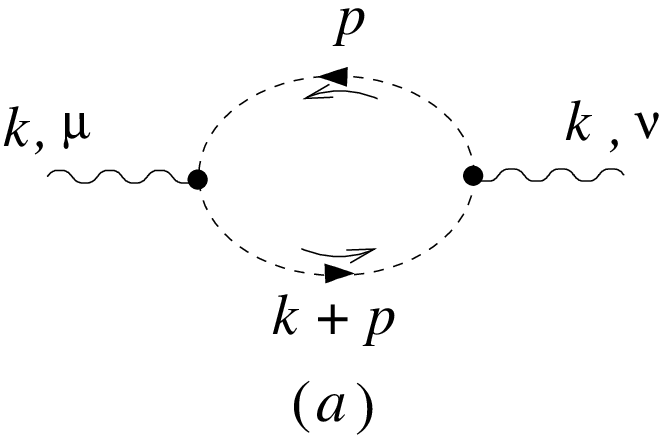}\end{array}&
\begin{array}{c}\includegraphics[scale=0.45]{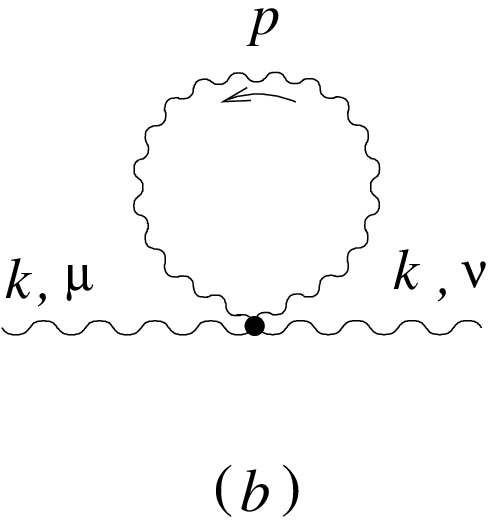}\end{array}
\\  
\begin{array}{c}\includegraphics[scale=0.45]{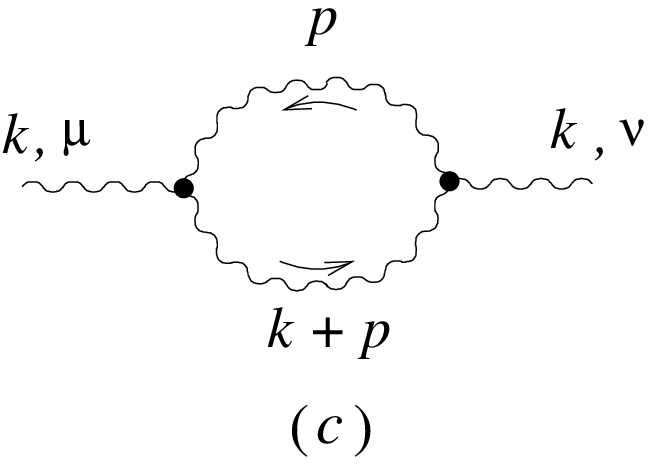}\end{array}&
\begin{array}{c}\includegraphics[scale=0.45]{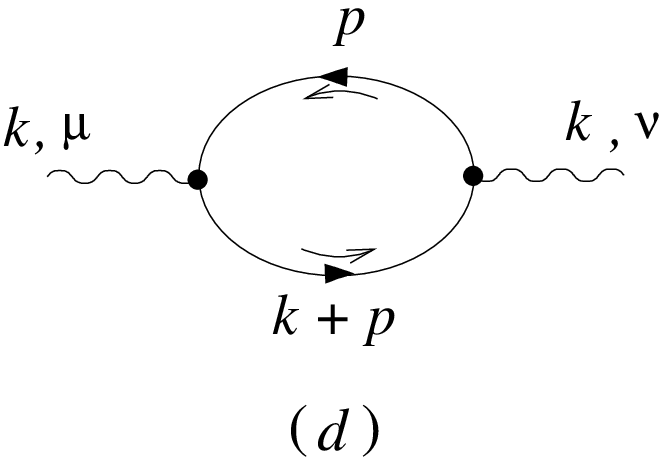}\end{array}
\end{array}
\]
\caption{
One-loop diagrams which contribute to the photon self-energy in NCQED.
}\label{ncloop}
\end{figure}
\end{center}

The diagrams which contribute to $\Pi^{\mu\nu}$ are shown in the
figure \ref{ncloop}. The first three diagrams, involving only bosonic loops,
have been computed in \cite{Brandt:2006bf}.
In the limit of large values of $\tau$, or, equivalently setting
$k_4=|\vec k|=0$ except inside the trigonometric factors, we have
obtained
\bea\label{1selfhighT}
\Pi^{\rm boson}_{\mu\nu}(k) & = &  \frac{(4 e)^2\, T^2}{2}
\int \frac{{\rm d}^{3} p}{(2\pi)^{3}} \sin^2\left(\frac{\tilde k\cdot p}{2} \right)  
\nonumber \\ & &
\!\!\!\!\!\!\!\!\!\!\!\! \times
\displaystyle{\sum_{n=-\infty}^{\infty} \left.\tilde\Pi_{\mu\nu}(p,k)\right|_{p4=2n\pi}}
;\; \tilde k_i \equiv k_j\theta_{ij} ,
\eea
where
\be\label{1tildePi}
\tilde\Pi_{\mu\nu}(p,k) =
\left[\frac{\eta_{\mu\nu}}{p^2}-\frac{2 p_\mu p_\nu}{p^4} \right].
\ee

In order to compute the sum and the integral in \eqref{1selfhighT}
we have employed the following decomposition
\bea\label{1pi3}
\Pi_{\mu\nu}^{\rm boson} & = & \Pi_{00}^{\rm boson}\,u_\mu u_\nu 
+\Pi_{\rm nc}^{\rm boson}\, \frac{\tilde k_\mu \tilde k_\nu}{{\tilde k}^2}
+ \Pi_{11}^{\rm boson}\,\frac{k_\mu k_\nu}{k^2}
\nonumber \\
& & 
\!\!\!\!\!\!\!\!\!\!\! +
\Pi_{22}^{\rm boson}\,\left(\eta_{\mu\nu} - u_\mu u_\nu - 
                                   \frac{k_\mu k_\nu}{k^2} 
 - \frac{\tilde k_\mu \tilde k_\nu}{{\tilde k}^2}\right),
\eea
where $u_\mu$ is the heat bath four velocity and $\tilde k_\mu \equiv k^\sigma \theta_{\sigma\mu}$.
Substituting \eqref{1selfhighT} in the left hand side of \eqref{1pi3}
and contracting the resulting equation with $u^\mu u^\nu$, 
$\frac{\tilde k^\mu \tilde k^\nu}{{\tilde k}^2}$, 
$\frac{k^\mu  k^\nu}{k^2}$ and $\left(\eta^{\mu\nu} - u^\mu u^\nu - 
                                   \frac{k^\mu k^\nu}{k^2} 
 - \frac{\tilde k^\mu \tilde k^\nu}{{\tilde k}^2}\right)$ 
we can express the four structure functions in terms of 
Matsubara sums and integrations over ${\rm d}^3p$, which can be done,
yielding the following results for the thermal part
\begin{subequations}\label{boson22}
\bea\label{1Af1}
\Pi^{\rm boson}_{00}(\tilde k) = \frac 1 2 \frac{(4 e)^2\,T^2}{(2\pi)^2}
\left(\frac{\pi^2}{6}+\frac{\pi}{2\tilde k}{\coth}(\pi\tilde k)
\right. \nonumber \\ \left.
-\frac{\pi^2}{2}{\coth}^2(\pi\tilde k)\right),
\eea 
\bea\label{1Df1}
\Pi^{\rm boson}_{\rm nc}(\tilde k) = - \frac 1 2 \frac{(4 e)^2\,T^2}{(2\pi)^2}
\left(\frac{\pi^2}{2}-\frac{\pi}{2\tilde k}{\coth}(\pi\tilde k)
\right. \nonumber \\ \left.
-\frac{\pi^2}{2}{\coth}^2(\pi\tilde k)  + \frac{1}{{\tilde k}^2}\right)
\eea
\end{subequations}
and $\Pi_{11}^{\rm boson}=\Pi_{22}^{\rm boson}=0$.

Let us now apply the same technique to the computation of the last
diagram in Fig. \ref{ncloop}. In fact, this diagram represents the sum
of the adjoint and the fundamental fermion loops. The contribution of
the fundamental fermion loop is well known and gives the following result in the high temperature limit
\be\label{00fund}
\Pi^{\rm fund}_{\mu\nu} = -\frac{e^2 T^2}{3} u_\mu u_\nu.
\ee
As we have already anticipated this contribution of fermions in the
fundamental representation does not modify the structure function $\Pi_{\rm nc}$;
it just gives the usual Debye screening mass of QED \cite{kapusta:book89}.
The contribution of the adjoint fermion can be readily obtained from
the Feynman rules generated by the first term in the action
\eqref{lagraferm}. Proceeding in the same way as in the case of the bosonic contribution,
we consider the limit of large $\tau$, which amounts to set
$k_4=|\vec k|=0$ except inside the trigonometric factor. This leads to
(reminding that the momenta have been scaled as $p\rightarrow p T$)
\bea\label{1selfhighT1}
\Pi^{\rm adj}_{\mu\nu}(k) = 4 e^2 T^2 \int \frac{{\rm d}^{3} p}{(2\pi)^{3}}   
\sin^2\left(\frac{\tilde k\cdot p}{2} \right)
\nonumber \\ \times
\sum_{n=-\infty}^{\infty}
\,{\rm tr}\left[\gamma_\mu\frac{p\!\!\!\slash}{p^2}\gamma_\nu
\frac{p\!\!\!\slash}{p^2}
\right]_{p_4=(2n+1)\pi}. 
\eea
Using 
${\rm tr}\gamma_\mu\gamma_\alpha\gamma_\nu\gamma_\beta = 4\left(
\eta_{\mu\alpha}\eta_{\nu\beta}+\eta_{\mu\beta}\eta_{\nu\alpha}-\eta_{\mu\nu}\eta_{\alpha\beta}\right)$,
we obtain
\bea\label{1selfhighTferm}
\Pi^{\rm adj}_{\mu\nu}(k) & = & -(4\,e)^2\, T^2
\int \frac{{\rm d}^{3} p}{(2\pi)^{3}}  \sin^2\left(\frac{\tilde k\cdot p}{2} \right)
\nonumber \\ &\times &
\sum_{n=-\infty}^{\infty} \left.\tilde\Pi_{\mu\nu}(p,k)\right|_{p4=(2n+1)\pi},
\eea
where $\tilde\Pi_{\mu\nu}$ is given by \eqref{1tildePi}.
It is remarkable that the result for the fermionic integrand 
in Eq. \eqref{1selfhighTferm} coincides, up to a factor, with the bosonic result in
Eq. \eqref{1tildePi}. 

Our explicit calculation in \cite{Brandt:2006bf} shows that there 
are two kinds of sums in terms of which the structure
functions in \eqref{1pi3} can be expressed, which are $\sum 1/p^2$ and
$\sum 1/p^4$. The thermal part of these sums can be expressed in terms of the
Bose-Einstein distributions $N_B$ for $p_4=2 n \pi$. On the other hand, in the present
fermionic case, with $p_4=(2n+1)\pi$, the Matsubara sums yield a
result which is simply related to the bosonic result in
\cite{Brandt:2006bf} by the replacement
$N_B\rightarrow - N_F$,  where $N_F$ is the Fermi-Dirac distribution.

Proceeding similarly to the bosonic case, we can now express 
$\Pi^{\rm adj}_{\mu\nu}(k)$ in terms of four structure functions,
using the same tensor basis as in Eq. \eqref{1pi3}. Then, the
structure functions can be explicitly computed performing the elementary
angular integrations as well as the slightly more involved 
integrations over $|\vec p|$ \cite{gradshteyn}.
As a result of these straightforward steps we obtain
\begin{subequations}\label{fermfinal}
\bea
\Pi_{00}^{\rm adj} &=& \frac{(4e)^2\,T^2}{(2\pi)^2}\left[
F(\tilde k) - G(\tilde k) - \frac{\pi^2}{6} \right]
\eea
\bea
\Pi_{{\rm nc}}^{\rm adj} = \frac{(4e)^2\,T^2}{(2\pi)^2}\left[
F(\tilde k) + G(\tilde k) +\frac{1}{\tilde k^2} \right]
\eea
\end{subequations}
and $\Pi_{11}^{\rm adj}=\Pi_{22}^{\rm adj}=0$, where, for notational convenience
we have introduced
\be
F(\tilde k) \equiv \frac{\pi}{4\tilde k}\left[
\tanh\left(\frac{\pi\tilde k}{2}\right)-\coth\left(\frac{\pi\tilde k}{2}\right)
\right] 
\ee
and
\be
G(\tilde k) \equiv \frac{\pi^2}{8} 
\left[\tanh^2\left(\frac{\pi\tilde k}{2}\right)-\coth^2\left(\frac{\pi\tilde k}{2}\right)
\right] 
\ee

The full result for the relevant part of the photon self-energy can
now be obtained adding the expressions in Eqs. \eqref{fermfinal} with the
corresponding ones in \eqref{boson22}. Using the notation 
$\Pi_{00} = \Pi_{00}^{\rm boson} + \Pi_{00}^{\rm fund} + \Pi_{00}^{\rm adj}$ and 
$\Pi_{\rm nc} = \Pi_{\rm nc}^{\rm boson} +\Pi_{\rm nc}^{\rm adj}$ we
can write the full expression for the photon self-energy as
\be\label{1pi3tot}
\Pi_{\mu\nu} = \Pi_{00}\,u_\mu u_\nu 
+\Pi_{\rm nc}\, \frac{\tilde k_\mu \tilde k_\nu}{{\tilde k}^2}.
\ee

\section{The Free Energy}

We have now all the ingredients to compute the free-energy in the
limit of large $\tau$. Taking into account the orthonormality and idempotency of the
quantities $u_\mu u_\nu$ and $\frac{\tilde k_\mu \tilde k_\nu}{{\tilde k}^2}$
and inserting Eq. \eqref{1pi3tot} into Eq. \eqref{eq1} 
(using $\sum_{n=2}^{\infty} {x^n}/{n} = -x -\log(1-x)$ in order to
perform the sum of all the rings), we find
\begin{eqnarray}\label{1tudo5a}
\Omega^{\rm r}(T,\theta) =
\frac{1}{2} 
\frac{T^4}{(2\pi\tau )^3}\int{\rm d}^3 k \!\!\!\! & & \left[
\frac{(e\tau)^2\, \bar \Pi_{00}(\tilde k) }{|\vec k|^2} 
\right. \nonumber \\
& & \!\!\!\!\!\!\!\!\!\!\!\!\!\!\!\!  
  +\left. \log\left(1-\frac{(e\tau)^2\, \bar \Pi_{00}(\tilde k) }{|\vec k|^2}\right) 
    \right . 
\nonumber \\ \qquad \;\;\;\;\;\;\;\;\;\;\;\;\;\;\;\;\;\;\;\;\;\;\;  
& & \left.
\!\!\!\!\!\!\!\!\!\!\!\!\!\!\!\!   
+ \frac{(e\tau)^2\, \bar\Pi_{\rm nc}(\tilde k) }{|\vec k|^2} 
\right. \\
& & \left. \!\!\!\!\!\!\!\!\!\!\!\!\!\!\!\!  
+ \log\left(1-\frac{(e\tau)^2\, \bar\Pi_{\rm nc}(\tilde k) }{|\vec k|^2}\right) 
\right]\nonumber ,
\end{eqnarray}
where we have introduced the quantities
$\bar \Pi_{00}(\tilde k)\equiv \Pi_{00}(\tilde k)/(e T)^2$ and
$\bar \Pi_{\rm nc}(\tilde k)\equiv \Pi_{\rm nc}(\tilde k)/(e
T)^2$.
This expression constitutes a contribution to the free-energy which is
non-analytic in the coupling constant, since it is not proportional to a simple power
of $e^2$.

There are many interesting physical consequences which can be drawn from 
Eq. \eqref{1tudo5a}. Usually, in the case of commutative theories, the 
components of the static self-energy are constants independent of the momentum $|\vec k|$.
For instance, even in the case of thermal gravity \cite{Gross:1982cv,Rebhan:1990yr},
the components of the graviton self-energy, in the high temperature
limit, are constants (thermal masses) which can be positive or negative. 
When these constants have a positive sign, the
free-energy picks up an imaginary part from the logarithm,
which can be interpreted as an instability \cite{Affleck:1980ac}.
In the present case of the TNCQED, the self-energy is a function of
the momentum $|\vec k|$ and also of the direction of the vector 
$\theta_i \equiv \frac 1 2 \epsilon_{ijk}\, \theta_{jk}$, except in
the case of the contribution of the fundamental fermion. 
Therefore, it is important to investigate the behavior of 
$\bar \Pi_{00}(\tilde k)$ and  $\bar \Pi_{\rm nc}(\tilde k)$ 
($\tilde k = |\vec k| \sin\alpha$ were $\alpha$ is the angle between
$\vec k$ and $\vec \theta$). Let us first obtain the asymptotic
behavior for small and large $\tilde k$. The Taylor expansion of expressions
\eqref{boson22} and \eqref{fermfinal} for small $\tilde k$ gives
\begin{subequations}\label{boson23}
\bea\label{1Af2}
\lim_{\tilde k\rightarrow 0} 
\bar \Pi^{\rm boson}_{00}(\tilde k) = 
-\frac{4\pi^2}{45}\tilde k^2 + \frac{4\pi^4}{315}\tilde k^4 + \cdots,
\eea
\bea\label{1Df2}
\lim_{\tilde k\rightarrow 0} 
\bar \Pi^{\rm boson}_{\rm nc}(\tilde k) = 
\frac{2\pi^2}{45}\tilde k^2 - \frac{8\pi^4}{945}\tilde k^4 + \cdots, 
\eea
\end{subequations}
\begin{subequations}\label{ferm23}
\bea\label{1Af3}
\lim_{\tilde k\rightarrow 0} 
\bar \Pi^{\rm adj}_{00}(\tilde k) = 
-\frac{7\pi^2}{45}\tilde k^2 + \frac{31\pi^4}{1260}\tilde k^4 + \cdots, 
\eea 
\bea\label{1Df3}
\lim_{\tilde k\rightarrow 0} 
\bar \Pi^{\rm adj}_{\rm nc}(\tilde k) = 
\frac{7\pi^2}{90}\tilde k^2 - \frac{31\pi^4}{1890}\tilde k^4 + \cdots.
\eea
\end{subequations}
For large $\tilde k$ these functions behave as
\begin{subequations}\label{largekt}
\bea\label{pi00b}
\lim_{\tilde k\rightarrow\infty} \bar \Pi_{00}^{\rm boson}(\tilde k) = -\frac{2}{3},
\eea
\bea\label{pincb}
\lim_{\tilde k\rightarrow\infty} \bar \Pi_{\rm nc}^{\rm boson}(\tilde k) = 0,
\eea
\end{subequations}
\begin{subequations}\label{largekt2}
\bea\label{pi00ad}
\lim_{\tilde k\rightarrow\infty} \bar \Pi_{00}^{\rm adj}(\tilde k) = -\frac{2}{3},
\eea
\bea\label{pincad}
\lim_{\tilde k\rightarrow\infty} \bar \Pi_{\rm nc}^{\rm adj}(\tilde k) = 0.
\eea
\end{subequations}
We may consider $\tilde k\rightarrow\infty$ as the limit when the two
opposite charges in the dipole are very far apart from each other.
This can be seen expressing $\tilde k$  in terms of the original dimensionfull
momenta so that 
$\tilde k \rightarrow |\vec k| \tau/T \sin\alpha=\theta T |\vec k| \sin\alpha$
and considering the limit $\theta T\rightarrow\infty$. Both the
adjoint fermion as well as the noncommutative photon can be pictured
as an extended dipole of effective length 
$L\sim \theta T$  \cite{Landsteiner:2001ky}.
In this case, loop contributions of photons and adjoint fermions
reduce to the sum of two fundamental fermion loops, corresponding to each charge in the dipole.
Since the self-energy is proportional to $e^2$, the two contributions add
to twice the result obtained for the fundamental fermion loop. This
interesting physical picture is indeed verified when we compare 
Eq. \eqref{00fund} with  Eqs. \eqref{pi00b} and \eqref{pi00ad}.

\begin{figure}[h!]
\begin{center}
\[
\begin{array}{c}
\begin{array}{c}\includegraphics[scale=0.3]{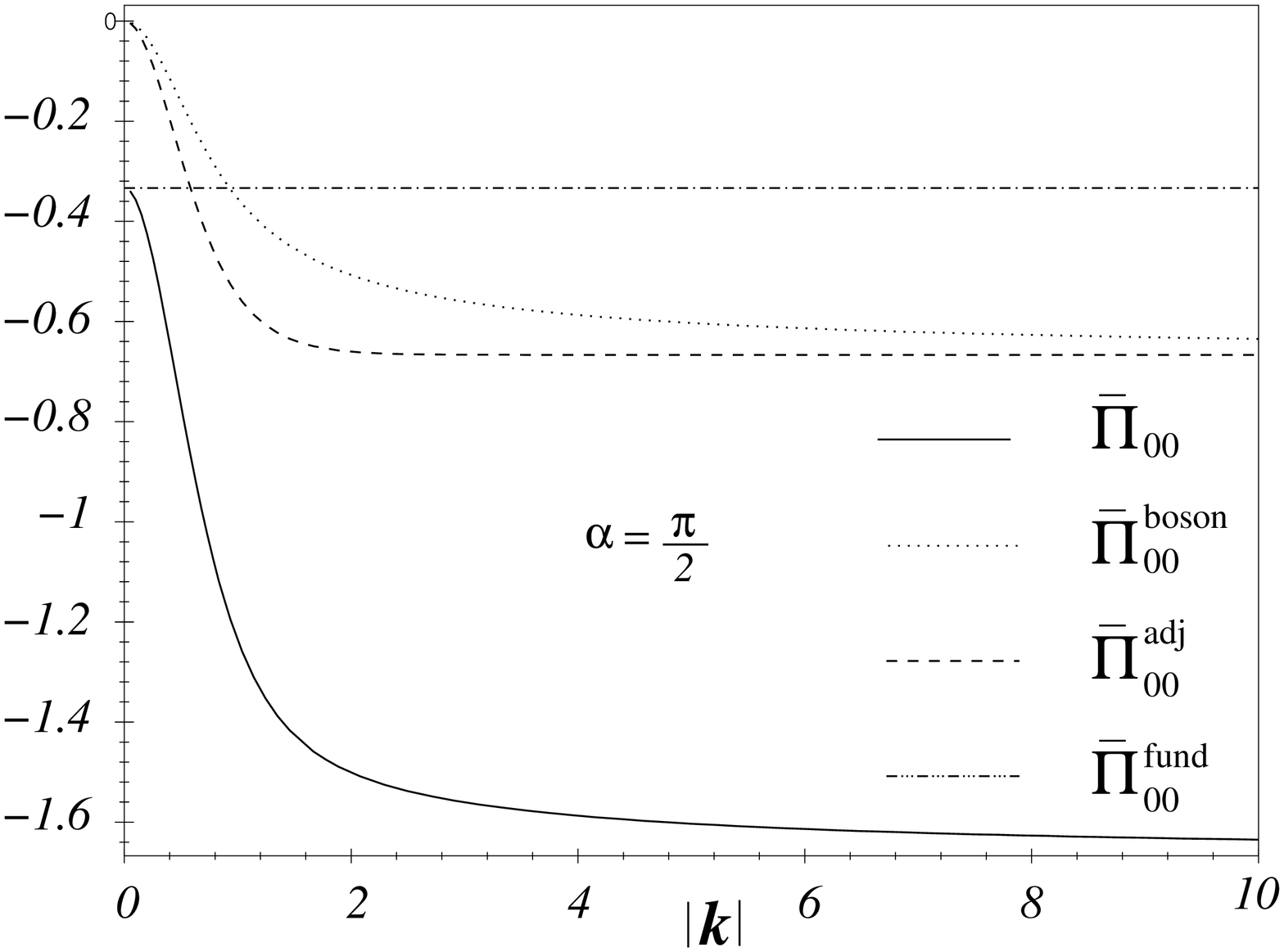}\end{array} \\
(a) \\ \\
\begin{array}{c}\includegraphics[scale=0.3]{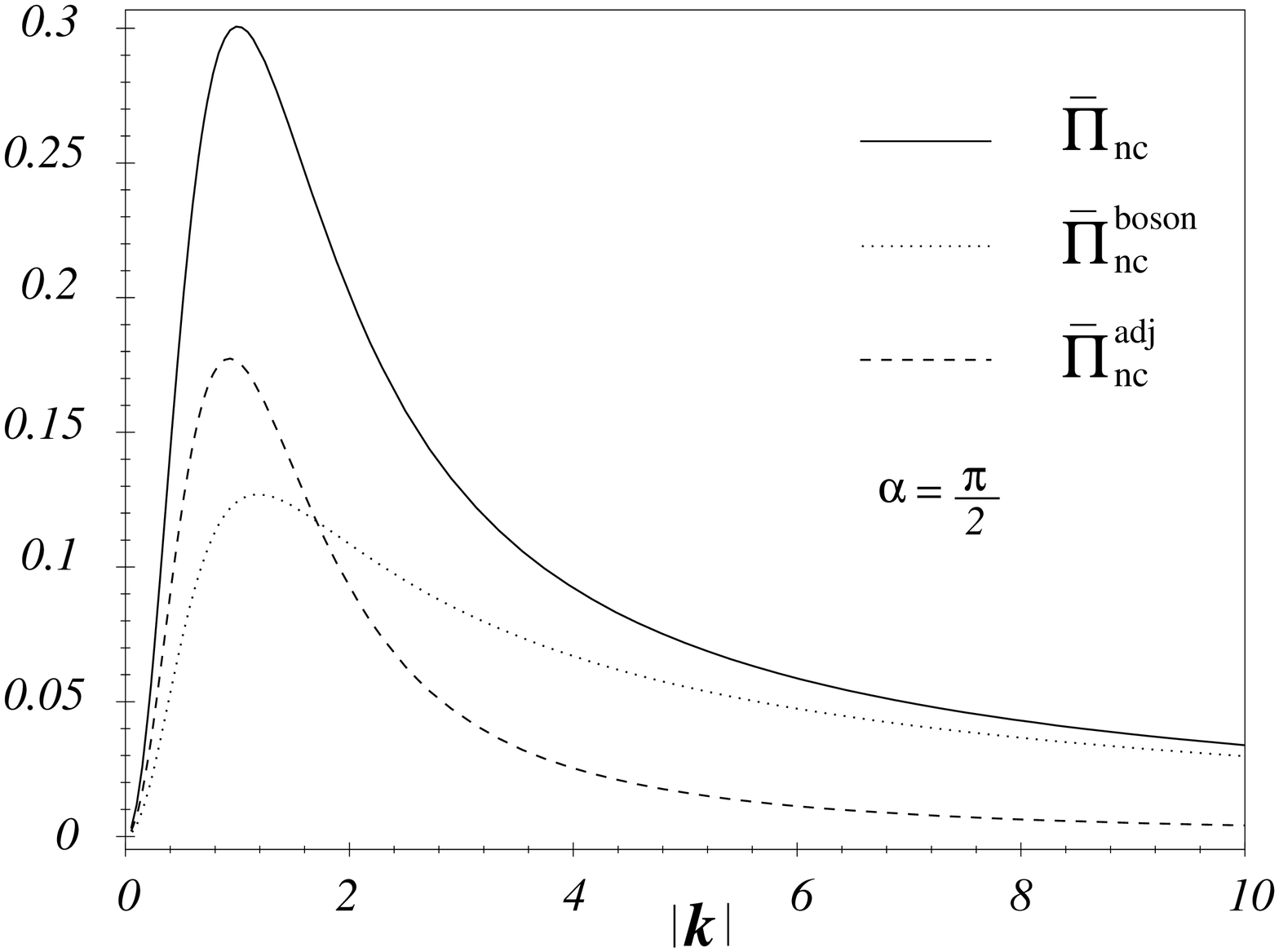}\end{array} \\
(b)
\end{array}
\]
\end{center}
\caption{The longitudinal (a) and transverse (b) modes of the static photon
  self-energy in NCQED.}\label{1AD1}
\end{figure}

Adding all the contributions (including also the contribution
from Eq. \eqref{00fund} such that 
$\bar \Pi^{\rm fund}_{00} \equiv \Pi^{\rm fund}_{00}/(e^2T^2)  = -\frac{1}{3}$), we obtain 
\begin{subequations}\label{tot111}
\bea
\lim_{\tilde k\rightarrow 0} 
\bar \Pi_{00}(\tilde k) = -\frac{1}{3}  -
\frac{11 \pi^2}{45}\tilde k^2 + \frac{47 \pi^4}{1260}\tilde k^4 + \cdots,
\eea
\bea
\lim_{\tilde k\rightarrow\infty} \bar \Pi_{00}(\tilde k) = -\frac{5}{3},
\eea
\end{subequations}
\begin{subequations}\label{tot222}
\bea\label{tot222a}
\lim_{\tilde k\rightarrow 0} 
\bar \Pi_{\rm nc}(\tilde k) = 
\frac{11 \pi^2}{90}\tilde k^2 - \frac{47 \pi^4}{1890}\tilde k^4 + \cdots,
\eea
\bea
\lim_{\tilde k\rightarrow\infty} \bar \Pi_{\rm nc}(\tilde k) = 0.
\eea
\end{subequations}
We also show in the figure \ref{1AD1} the plots of the functions $\bar\Pi_{00}$ and $\bar\Pi_{\rm nc}$ (full
lines) as well as their  bosonic and fermionic components (doted and dashed lines). 

From the behavior of $\bar \Pi_{\rm nc}$ and $\bar \Pi_{00}$ we can now
investigate the thermodynamic stability as follows.
Figure \ref{1AD1}a shows that the contribution of the bosonic sector
to $\bar \Pi_{00}$ does not induce an imaginary part to
the free-energy in \eqref{1tudo5a}, because
$\bar\Pi_{00}^{\rm boson}$ (doted line) is always negative. 
Similarly, the contribution to $\bar\Pi_{00}$ from the
adjoint fermions (dashed line) is also negative.
Hence, the longitudinal $00$ mode is always stable.

In the case of the transverse mode, figure \ref{1AD1}b 
shows that the bosonic sector (doted line) is positive, yielding the
instability already discussed in ref. \cite{Brandt:2006bf}. The
adjoint fermion contribution is also positive.
Therefore, the sums of
the fermionic and the bosonic contributions produce a larger 
positive result for the noncommutative transverse mode,
which enhances the instability.

Finally, we can investigate how the critical value of the temperature
given in \eqref{tc1} is modified by the inclusion of fermions. This
can be done by realizing that the second logarithm in \eqref{1tudo5a}
has an imaginary part when
\be
(e\tau)^2 > \frac{k^2}{\bar\Pi_{\rm nc}(\tilde k)} \ge 
\frac{\tilde k^2}{\bar\Pi_{\rm nc}(\tilde k)} .
\ee
This condition can be solved analytically and leads to the
following critical temperature
\be\label{tc2}
T_c = \sqrt{\frac{1}{e\theta} \frac{3}{\pi}
\sqrt{\frac{10}{11}}} \approx .954\, \sqrt{\frac{1}{e\theta}}
\ee
which is associated with the noncommutative transverse mode.
Comparing with \eqref{tc1} we conclude that the inclusion of the 
adjoint fermion has decreased the critical temperature.

\section{Discussion}
In conclusion, our results reveal that the possible excitation of
degrees of freedom of the adjoint fermion renders the system more
unstable, due to a decrease in the value of the critical temperature,
when compared to the one obtained in \cite{Brandt:2006bf}. This
happens because the thermal contribution of the adjoint fermion is
qualitatively similar to the one of the (noncommutative) photon. 
This behavior might be expected since one may picture all such
particles as having a likewise dipole structure
with zero net electric charge
\cite{Hayakawa:1999yt}. Consequently, the total contribution from the
photon and the adjoint fermion to the transverse noncommutative 
magnetic mode makes $\Pi_{\rm nc}$ even larger, which enhances the instability of
the system.

One may understand this instability, if we consider the noncommutative
thermal magnetic mass defined in a self-consistent way as
\be\label{d01}
m_{\rm nc}^2 = - \left.\Pi_{\rm nc}(k_0=0,\vec k)\right |_{k^2=m_{\rm nc}^2}.
\ee
Proceeding similarly as in \cite{Brandt:2006bf}, we find that this
equation admits negative solutions for $m_{\rm nc}^2$. This occurs in
the region $e\tau > e\tau_c$, which is where the argument of the
second logarithm in Eq. \eqref{1tudo5a} becomes negative. When $e\tau$ is
close to its critical value, we can obtain an analytic solution of
Eq. \eqref{d01}, using for $\bar\Pi_{\rm nc}$ the expression given by
Eq. \eqref{tot222a}. After a simple calculation, we find that the
noncommutative thermal magnetic mass can be written in this region in the form
\be
m_{\rm nc}^2 = \frac{847}{705}e^2\left({T_c}^2-T^2\right),
\ee
where $T_c$ is given by Eq. \eqref{tc2}. This solution shows explicitly
that $m_{\rm nc}^2$ becomes negative when $T>T_c$.

It is interesting to note that a negative value of the squared
magnetic mass is reminiscent of the negative squared Jeans mass 
$M^2\sim -G T^4$ which arises in quantum gravity 
\cite{Gross:1982cv,Rebhan:1990yr}. Such a mass leads to the appearance
of an imaginary part in the free energy, which indicates that the
system becomes unstable and may undergo a phase transition
\cite{Affleck:1980ac}.

\bigskip

\noindent{\bf Acknowledgment}

The authors would like to thank CNPq and FAPESP, Brazil, for financial support.



\end{document}